\documentclass[12pt,preprint,epsfig]{aastex}

\received{}

\revised{}

\accepted{}

\cpright{AAS}{2001}

\ccc{000}

\lefthead{Bianchi, Bohlin, Massey}
\righthead{ Ofpe/WN9 stars in M33 }

\slugcomment{Version of  \today}

\newcommand{\as}{\mbox{$^{\prime\prime}$}}
\newcommand{\ebv}{$E(B-V)$}

\newcommand{\Teff}{\mbox{$T_{\rm eff}$}}
\newcommand{\Lbol}{\mbox{$L_{\rm bol}$}}
\newcommand{\lbol}{\mbox{$L_{\rm bol}$}}

\newcommand{\Lx}{\mbox{{$L_x$}/$L_{\odot}$}}

\newcommand{\Msun}{\mbox{$M_{\odot}$}}

\newcommand{\Vinf}{\mbox{$V_{\infty}$}}

\newcommand{\kms}{km s$^{-1}$}
\newcommand{\Myr}{M$_{\odot}$ yr$^{-1}$}
\newcommand{\myr}{M$_{\odot}$ yr$^{-1}$}
\newcommand{\mdot}{$\dot M$}
\newcommand{\Mdot}{$\dot M$}

\newcommand{\dlam}{\mbox{$\lambda\lambda$}}
\begin{document}
\title{The Ofpe/WN9 stars in M~33
\footnote{Based on observations with the \it Hubble Space Telescope
\rm which is operated by AURA, Inc. under NASA contract NAS5-26555} }
\author{Luciana Bianchi} 
\affil{Center for Astrophysical Sciences,
The Johns Hopkins University,
239 Bloomberg Center for Physics \& Astronomy,
3400 N. Charles str., Baltimore, MD 21218, USA -
Email: bianchi@pha.jhu,edu}
\author{Ralph Bohlin} 
\affil{Space Telescope Science Institute, 3700 San Martin Dr., Baltimore, MD 21218}
\author{Philip Massey} 
\affil{Lowell Oservatory, 1400 W. Mars Hill Rd, Flagstaff, AZ 86001}

\begin{abstract}

We present HST/STIS Ultraviolet spectra of the six known Ofpe/WN9
stars (``slash stars'') in M~33. These stars were selected 
for showing the characteristics of the Ofpe/WN9 class from 
previous optical ground-based spectroscopy.  The UV spectra are rich in
wind lines, whose strength and terminal velocity vary greatly
among our target sample. We analyse the STIS spectra with non-LTE,
line blanketed, spherical models with hydrodynamics, computed with
the WM-basic code. We find  C to be  underabundant and N  overabundant,
respect to the solar values,
with a ratio (by mass) of C/N between 0.02 to 0.9 across the sample.
Some stars show very conspicuous wind lines (P Cygni profiles), while
two stars have extremely weak winds. The mass-loss rates thus vary
greatly across the sample. 
 The mass-loss rates of the hottest stars  are
lower than typical values of WNL stars, but higher than expected for
normal population I massive stars. There is indication that the mass-loss rates
 may be  variable in time. 
The C/N ratio, and the other
 physical parameters derived by the spectral modeling (\Teff , \Lbol, mass), 
are consistent with evolutionary calculations for objects
 with moderately high initial masses ($\approx$30-50\Msun),
 evolving towards the WNL stage through an enhanced
mass-loss phase. 
  \end{abstract}

\section{Introduction}
\label{sintro}

There is a general consensus about the Wolf-Rayet stars being formed from
O stars through high mass-loss phases, but many specific questions remain
still open.
The  Ofpe/WN9 stars (``slash stars")  
class, introduced by Walborn (1977, 1982a),
shows properties intermediate between those of Of and WN stars:
the emission is stronger than in the most extreme Of stars, but HeII
photospheric absorption indicates a less dense wind than seen in any
Wolf-Rayet stars (see Bohannan \& Walborn 1989). The implication is that
 Ofpe/WN9 stars are transition objects between the Of stars and the
Wolf-Rayet in the ``Conti scenario" (Conti 1976; Conti \& Bohannan 1989).
Based on the reduced surface  H fraction observed in these stars,
Pasquali et al (1997) suggest instead the sequence O
$\rightarrow$ Of $\rightarrow$ H-rich WNL $\rightarrow$ Ofpe/WN9,
for initial masses less than 100 M$_{\odot}$.
One of the prototypes of the Ofpe/WN9 stars, R127 (HDE 269858) underwent
an ``S Dor like'' eruption in 1982 (Walborn 1982b; Stahl \it et al. \rm 1983)
revealing Luminous Blue Variable (LBV) characteristics. 
Other evidence  suggests that at least some of the Ofpe/WN9 stars
  have progenitors that are considerably lower in mass than the WNLs; see
  Massey, DeGioia-Eastwood, \& Waterhouse (2001). 
LBV's irregular
episodes of extreme mass-loss may play an important role in the formation
of W-R stars by substantial shedding of the progenitor's outer layers
(e.g. Maeder \& Conti 1994).

Ten ``slash stars'' are known in the LMC (Bohannan \& Walborn 1989).
Seven  LMC Ofpe/WN9 stars were investigated in the 
optical (Nota et al. 1996) and  in the UV with 
Faint Object Spectrograph (FOS) spectra (Pasquali
et al. 1997), and with IR spectroscopy
 by Morris et al (1996). The spectral morphology 
confirms the transition  nature of these objects and their
resemblance to LBVs.  Crowther et al. (1977) also analysed optical
spectra of LMC Ofpe/WN9 stars and re-classified them as WNL. 
Krabbe et al. (1991) observed a number of HeI emission line stars
in the Galactic Centre, whose HeI line ratios and line-to-continuum
ratio are consistent with  Ofpe/WN9 stars.
An  Ofpe/WN9 star in NGC 300 was reported  by Bresolin et al. (2002). 

With follow-up, ground-based (KPNO, MMT) spectroscopy of about 400
UV-brightest stellar sources found in the \it Ultraviolet Imaging
Telescope \rm (UIT)  far- and near-UV images of M~33, 
 we discovered six Ofpe/WN9 stars
in this galaxy (Massey et al. 1996, hereafter MBHS). 
The ``slash stars'' are UV bright. In our study of the UIT
UV-brightest sources in M~33, these stars are relatively
well separated from other stars in the $ U - B $ vs. $ FUV - NUV $
color-color diagram. They lie near the LBV candidates. Their colors are
consistent with the fact that the Balmer jump is in \it emission \rm
in Ofpe/WN9 stars, which requires both high UV flux and an extended
atmosphere, consistent with the strong P Cygni profiles (e.g.
HeI $\lambda$ 4471)
characteristic of the Ofpe/WN9 class.

We present {\it Space Telescope Imaging Spectrograph }
(STIS) ultraviolet spectra of the six known Ofpe/WN9
stars in M~33. 
In section 2 details are given about the observations and the
reduction, in section 3 the STIS UV spectra are analysed with 
non-LTE, line blanketed, hydrodynamical, spherical model atmospheres to derive physical
stellar parameters.  The results are interpreted in terms of
stellar evolution by comparison with evolutionary model predictions (Section 4).

\section{The programs stars. Data and reduction}
\label{sdata}

\subsection{The program stars}
\label{s_stars}
In Bianchi's HST program GO8207, 
we obtained UV spectra of the six Ofpe/WN9 stars known in M33. The spectra
are shown in Figure \ref{f_spectra_all}. 
The objects were classified
as Ofpe/WN9  from ground-based (KPNO,
MMT) classification spectra of the 
UV-brightest stellar sources measured in the UIT far-UV and near-UV
 images (MBHS). The optical spectra, taken several years earlier (1993 - 1995), are
shown in Figure \ref{f_spectra_optical} for the purpose of discussion in 
the following sections. 
A seventh object (M33-UIT339) was included in our program  because of its
photometric properties (H$\alpha$ \ emission-line source ).
An additional  object, M33-UIT340, was
 included serendipitously  in the STIS long slit during the M33-UIT339 observations 
by chosing an appropriate orientation.
Its spectrum  has a UV flux level comparable  
to the primary target, but much stronger UV lines, in spite of the similar
optical colors.
Another serendipitous star was observed due to failure of the acquisition
on a repeated observation of M33-UIT349.

Data from our previous ground-based photometry 
are compiled in Table \ref{tdata}. In the table
we give the star
name, coordinates and V-mag, from the UIT sources list of MBHS.
Also, for the position of every star we computed the de-projected
galactocentric distance, using PA = 22$^{\circ}$ 
for the position angle of the semi-major axis of M~33, an inclination of the
disk  {\it i}=54$^{\circ}$ 
and  a systemic velocity for M33 of V$_{sys}$=-180 \kms\ ~(from Warner et al.
1973). Warner et al. (1973) also provide a model for the rotation of the 
M33 disk from HI measurements. We computed the expected disk velocity,
projected along the line of sight, at the position of each star. 
Because the targets are likely to lie in the M~33 disk, we used this
velocity to shift the spectra to a rest-wavelength frame at the star. 
The position of the stellar lines indicates that this velocity 
correction is appropriate in all cases, although the difference
from just using the M33 systemic velocity is not really appreciable at 
the resolution of our spectra.  Both the galactocentric distance
and the recessional velocity are also given in Table \ref{tdata}.  
Finally, the STIS datasets  are listed.

\subsection{STIS spectroscopy and data reduction}
\label{s_redu}

The observations  were completed  over a long time span, because the HST
entered a safe mode event just after our 
program (GO8207) began, and most of
the observations were delayed 
because we missed the
window for the desired orientation constraints. 
To assure homogeneous data quality and consistent flux calibration, we 
used the most recent on-the-fly calibration for all of the data after
the program was completed. 

\begin{figure}
\plotone{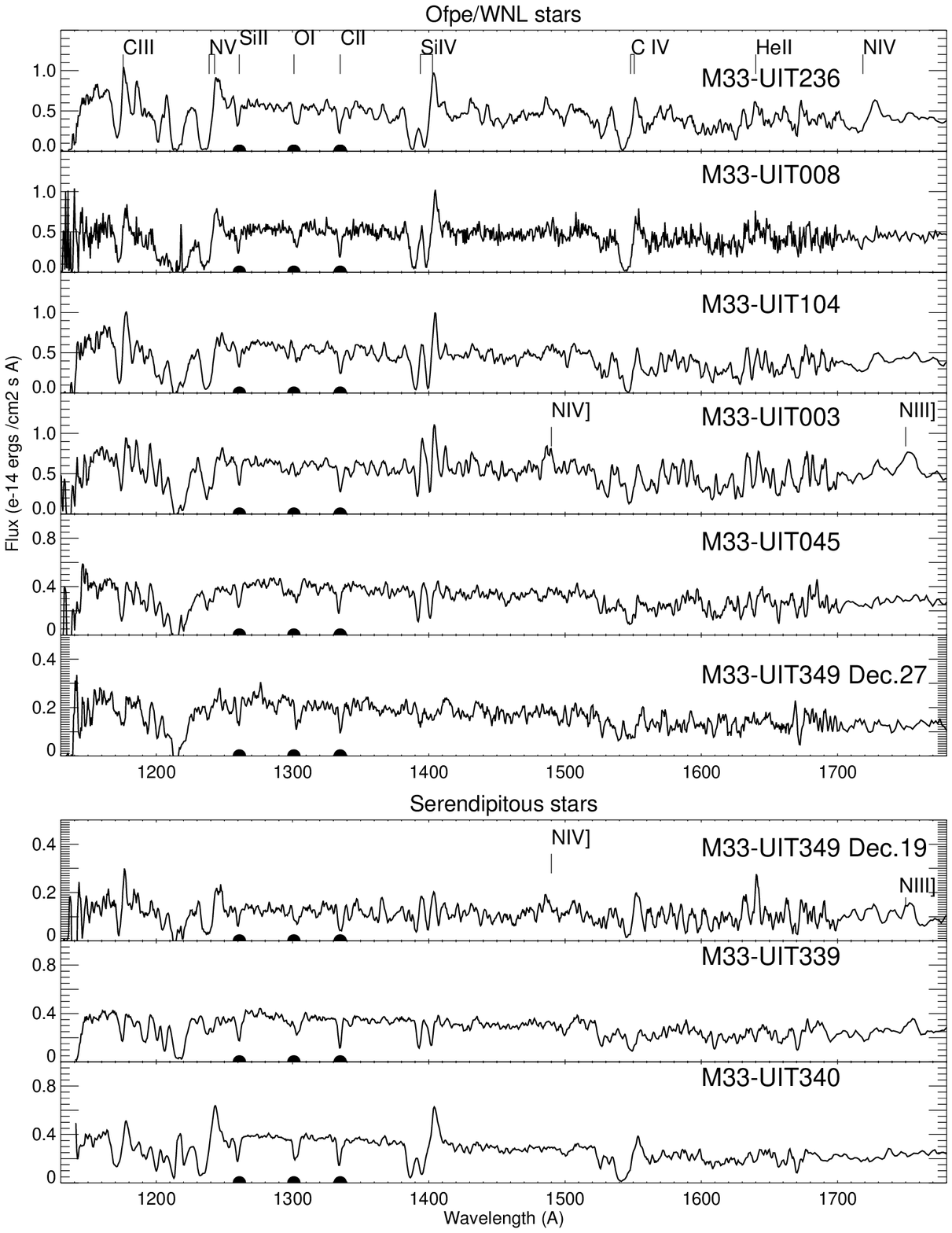} 
\caption{The UV spectra in the short wavelength range, which contains the lines
of interest: CIII, CIV, NIV, NY, SiIV, HeII.
 The Ofpe/WN9 program stars are
arranged by progressively decreasing 
strength of the UV wind lines. The spectra of the serendipitous stars
(see text) are shown at the bottom. 
The strongest interstellar lines in this range,
\ion{Si}{2}$\lambda$1260, \ion{O}{1}$\lambda$1301, and \ion{C}{2}$\lambda$1335,
are marked with a filled bubble at the bottom of the spectra. 
The strongest stellar features are also marked on the top spectrum.
\label{f_spectra_all} }
\end{figure}

For each star the same observing pattern was repeated: after target acquisition,
a 24 min. long  exposure  was obtained with the STIS G230L grating
(range 1570-3180 \AA, scale 1.58\AA / pixel), followed by an 8 min.
exposure in the same orbit with the grating G140L (range 1150-1730\AA ,
scale 0.60\AA /pixel). The whole subsequent orbit was devoted to a long
(57 min) G140L exposure. The total exposure time was therefore
3300sec (G140L grating) and 1440sec (G230L grating) for every target.     
The STIS 0.2\arcsec\ x 52 \arcsec\  
slit was used.
For target M33-UIT349, which is fainter than the others in the sample, we
repeated this pattern twice, aiming at doubling the total time and
 achieve a S/N comparable to the other sample objects.  The two sets of
 observations  were taken about a week apart (Dec. 19 and Dec. 27, 2000).
The spectra are significantly different on the two dates, 
both in flux level and wind lines, 
as can be seen in Fig \ref{f_spectra_all}. 
After examining 
the STIS CCD50
acquisition images, we discovered that the intended target (M33-UIT349)
was actually observed on Dec. 27, while a nearby star 
was centered in the aperture on Dec. 19, during
the fine centering stage, 
in spite the right target was exactly centered (on the same pixel) 
after  coarse aquisition on both dates, as shown in Figure \ref{f_acq_349}.
We measured photometry  on both dates,
on the STIS CCD acquisition images:
the two stars are equally bright in the CCD50 filter. 
Additionally, the target (M33-UIT349) appears slightly elongated in the
acquisition images (Figure  \ref{f_acq_349}),
but it is not clearly resolved into
multiple sources.   A cut of the long slit 2D spectra along the
spatial direction, by summing several columns across the wavelength dispersion
(Figure \ref{f_ycut}), 
shows a hint of a faint extension in  M33-UIT349 spectrum (Dec. 27,
datasets O5CO07010,20,30), although the two peaks, or asymmetric
intensity profile,  may just be an effect of the low countrate.
 The STIS MAMA spatial scale is 0.025''/pxl,
and the width of the entire spectrum (see Figure \ref{f_ycut}) 
is about 10 pixels or .25'', with
the two brightest components less than .1'' apart. Therefore it appears
that, although the object is not resolved into multiple stars in
 the STIS CCD, which has a  factor of two lower
resolution than the MAMA, there may be  some [marginal] extra light into the
STIS slit from other fainter sources.
The spectra are  centered in the y-direction (perpendicular
to the dispersion) in the nominal position. 
Just based on the elongation seen on the finding chart, and
the two apparent peaks in the spectrum intensity profile, 
we can speculate: could  
  the object  be a  binary (multiple?) system, or is there a
 chance superposition of a faint source?  A separation
of $\leq$0.1'' at the distance of M33 (840,000 pc, Magrini et al. 2000) would correspond to 
$\leq$ 0.4pc, extremely large for a bound binary. 
An additional source seem to be excluded by  the quantitative modeling
of the spectra.

We will call the serendipitous star which was in the slit on Dec. 19,
 M33-UIT349-B hereafter. In  the acquisition
image, it is just 1.33$''$ east of the target, thus the coordinates given 
in Table \ref{tdata} are relative to the ground-based coordinates of the source
(+0.10 seconds in R.A.). Most likely, the ground-based coordinates of M33-UIT349
were an intermediate position between the two sources, whose
separation of 1.3$''$ is not resolved in the ground-based image,
and is comparable to the coordinate accuracy. The two stars can uniquely
be identified from their relative positions and our finding chart
(Figure \ref{f_acq_349}).

For every target star we examined the two G140L exposures taken in subsequent
orbits, and found that they
perfectly match in absolute flux and the strong line features reproduce
very well - in spite of the very different S/N due to the different exposure
time. Also, we verified that the 
G230L and G140L fluxes match in the wavelength region of overlap. 
With the exception of M33-UIT339 (see later), we then combined the two G140L
 exposures, taking into account the different exposure times, to
obtain the maximum possible  total S/N over the range. 
The STIS flux calibration 
is expected to be accurate for sources in the broad 2\as ~ wide
slit to 2\% (Bohlin 2000; Bohlin, Dickinson, \& Calzetti 2001).
For the 0.2\as ~ slit used in this program, photometric precision of
one wavelength with respect to a distant wavelength is 4.5\%rms (Bohlin
\& Hartig 1998). There are no known non-linearities in the response of 
the STIS MAMA detectors. However, additional uncertainties arise from
the Poisson counting statistics and from possible errors in background
subtraction for the faintest sources.

An additional target, M33-UIT339, was included in our program GO8207
as it appeared to be an H$\alpha$ emission-line source, however it was not 
subsequently classified as an Ofpe/WN9 star.
With an appropriate orientation, the 
 STIS long slit centered on M33-UIT339 included also M33-UIT340,
separated by about 4$\arcsec$ . 
The  pipeline automatic extractions for the two G140L spectra,
taken in subsequent orbits, are centered on the two differemt stars
respectively, therefore the extracted spectra from the archive should {\it not} be used. 
The automatic pipeline extraction
has centered the extraction slit on the target in one case 
(dataset O5CO03020), and on
the serendipitous star in the other case (dataset O5CO03030). Therefore 
we manually  re-extracted the spectra of both stars 
 in all three exposures (two G140L and one
G230L) and coadded them separately. 
While there is no acquisition image for the second orbit (only reacquisition
of previous guide stars was performed), we did obtaine a parallel  WFPC2 image with
the same filter (F170W) in each orbit, as well as with different
filters. Our F170W WFPC2 images with 0.1$\arcsec$ resolution 
show that there is no shift between the
two pointings  down to the pixel level.
The continuum level is the same over most of  
the G140L range, but the wind features are
completely different for the two stars (Figure \ref{f_spectra_all}).

The spectra of the serendipitous stars are shown in Figure  \ref{f_spectra_all}
for completeness, and for clarification of all aquisition and multiple-source
issues to potential archive users. They will be analyzed separately.

Because the targets were selected from ground-based photometry and
spectroscopy, we examined all  the STIS acquisition
images 
to check for multiplicity that might be
revealed at the HST STIS imaging resolution but not in the ground-based data.
All targets appear point-like. 

\section{Analysis of the UV spectra}
\label{sanalysis}

\subsection{UV line morphology}
\begin{figure}
\plotone{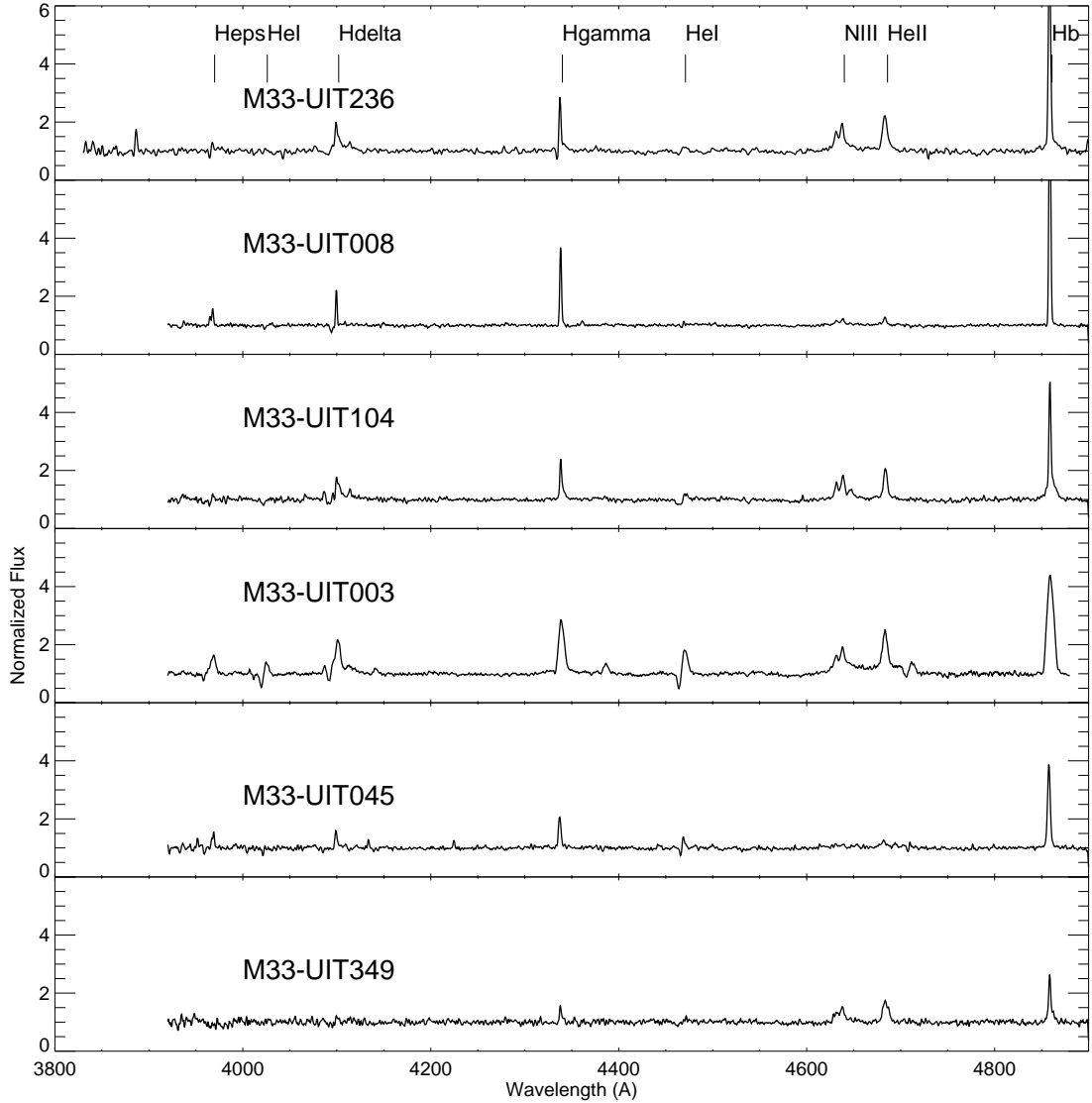} 
\caption{The optical spectra of the program stars (about 2\AA~ resolution, from MBHS), 
are shown arranged in the same order as the UV spectra in Fig. \ref{f_spectra_all}.
To facilitate appreciation of relative line strength, the spectra are normalized
to the continuum, and plotted on identical scale. The relative line
strength does not match the progression seen in the UV lines, suggesting
variability. 
\label{f_spectra_optical} }
\end{figure}

The short wavelength ($<$ 1750\AA) portion of the spectra for all the 
program stars and the serendipitous stars is shown in Fig. \ref{f_spectra_all}.  
This wavelength region is richest in lines from the stellar wind 
(and the interstellar medium). 
The main wind lines are
shown in detail in  Figure \ref{f_vinf}, where their extent in velocity
(largely varying across the sample) can be appreciated.
 The spectra are arranged from
top to bottom of the figures in a progression of decreasing strength of the UV lines
and decreasing wind velocity (width of the P Cygni absorption). 
Previous ground-based spectroscopy from MBHS
also showed a range of line strengths within the sample.
However, the progression is different from what the recent UV spectra
indicates. The optical spectra from MBHS are shown in Figure \ref{f_spectra_optical}. 
For example, M33-UIT008 was a weak-liner in the  optical spectra, but displays
some of the strongest UV features in the STIS spectra. Viceversa,
M33-UIT349 has fairly strong optical lines (Figure  \ref{f_spectra_optical}),
 but not in the  UV spectra. 
As another example, M33-UIT045 has  weak lines both in
the old optical spectra and in the UV spectra. 

The UV line strength progression
indicates a range of \Teff (luminosities) and mass-loss rates 
among the sample. It is difficult to relate the progression to 
 an evolutionary sequence, because  the relative line
strengths are different between the optical spectra ($\approx$ 1993 -1995)
and the recent (2000-2001) UV spectra, suggesting variability.
One stellar parameter which can plausibly
vary on very short time-scales is the mass-loss rate, if these objects
were single stars. Or are they evolved binaries? 
The quantitative spectral modeling described below is aimed at
claryfying the evolutionary status, by measuring the  stellar parameters.

\subsection{NTLE modeling with WM-basic}

We  performed detailed quantitative modeling of the spectra with
the WM-basic code (Pauldrach et al. 2001).
The code solves the wind hydrodynamics equations for specified parameters
and includes non-LTE metal line blanketing in spherical geometry. It also
allows us to include the effects of shocks (from clumps)  on the wind ionization
in the radiative transfer calculations, which are extremely important
for  a correct modeling of hot massive stars (e.g. Bianchi \& Garcia 2002, 2003).
Below we discuss the determination of the 
physical parameters for each individual
star and how the uncertanties were estimated.
More discussion on the uncertainties can be found in Section 4. 
The parameters from the best fit models are
listed in table \ref{tresults}. 
The process is quite complex, given the many physical parameters
which affect the emerging spectrum. 
The analysis of the first object is thus
described in  detail as a sample case. The effects of the physical
parameters on the emergent spectrum are similar for the other objects.
Initial guesses for the terminal velocities were taken by measuring
the maximum extent of the P Cygni profile absorptions, which largely
varies across the sample as can be appreciated (on an expanded scale, and independent
of any modeling) in Figure  \ref{f_vinf}.

\subsubsection{M33-UIT236}

The UV spectrum of this object indicates that it might be the hottest of 
the sample, and that it has a high mass-loss rate, because it has the strongest
UV wind lines. We ran a grid of WM-basic models, with initial parameters
covering a range  of values for  \Teff , luminosity and mass-loss rate 
inferred from a wider grid  computed for other purposes. 
We computed synthetic model spectra for both solar 
and LMC-type metallicities, and then we also
varied individual abundances of the CNO elements. 
In Table \ref{tresults} we give the resulting
stellar parameters and their uncertainty range, and here we discuss how
they were constrained.  

Before we discuss the strong P Cygni profiles that primarily constrain the 
main stellar parameters, we note that all models of the initial grid computed
for LMC-type metallicity (z=0.008) are ``flatter'' than the observed spectra
 in the some spectral regions which contain
 no single strong lines, but a multitude of fainter lines,  mainly from 
\ion{Fe}{3} and \ion{Fe}{4} transitions (e.g. from 1420 to 1500\AA).
In these regions, 
the emerging flux in the low (LMC-type) 
metallicity models is close to a pure continuum,
while the observed spectrum definitely shows 
 shallow but significant depressions. These can be reproduced 
better in models with solar metallicity ({\it z}=0.02). 
A precise determination of {\it z} is not possible. 
Therefore we adopted solar
metallicity in the subsequent computation of the finer grid which includes the
best fit adopted solutions, discussed hereafter.  
The result is not surprising given the metallicity gradient 
in M33 and that M33-UIT236 is relatively close to the galaxy
center. 
The model spectra are sensitive to this effect only for large
wind velocities. 

 The STIS spectrum of M33-UIT-236 (Figure \ref{f_spectra_all})
shows strong P Cygni profiles of both \ion{Si}{4} and \ion{N}{5}, as well
as a strong \ion{C}{4} $\lambda\lambda$ 1548,1551 doublet  and \ion{C}{3} multiplet
around 1174\AA. To reproduce the observed strength 
 of both \ion{N}{5} and \ion{Si}{4} consistently,
we need to take into account the X-rays produced by shocks in the wind
as a relevant ionization process. 
Because there is no prior information, to our knowledge, about soft X-ray
emission from these objects, we tried a range of values that would be 
appropriate for early type luminous stars as observed e.g. in our Galaxy
(e.g. Bianchi 1982a,b, Bianchi \& Garcia 2002a,b). 
We found that for  \Teff = 32000~K or lower, \Lx $\approx$ -7 is appropriate
to reproduce the strength and shape of the \ion{N}{5} doublet and the other
``hot'' transitions relative to the lines 
from lower ionization potential stages such as 
\ion{Si}{4} .   
A lower X-ray luminosity of \Lx=-7.5 is found to be adequate at 
 a higher \Teff (34,000K). 
The \ion{Si}{4} \dlam 1394,1403 
doublet, as well as \ion{N}{4} $\lambda$ 1718, is not very sensitive to the effect of 
shocks, but is instead very sensitive to \Teff (especially in the range 
34000 - 30000~K) and to the mass-loss rate. 
HeII $\lambda$ 1640 is very sensitive
to the mass-loss rate, in this temperature regime, thus cannot be used
to constrain the helium abundance.
The {\it  relative} strength 
of the \ion{C}{4} to \ion{C}{3} lines is again best reproduced by models
between \Teff = 32000~K and \Teff = 34000~K. 
However, both of  these C  lines 
are always  too strong for all parameter combinations
which can reproduce the overall spectral features, in the 
models with solar abundances.
We have therefore progressively decreased the C abundance
(and changed correspondingly the abundance of N and O as expected
in different evolutionary scenarios) 
 and finally achieved  a consistent fit of all the strongest wind lines
with  carbon  20 times less abundant than the solar value and
nitrogen 10 times overabundant. These values correspond to the abundances
predicted by  evolutionary models for e.g. a star with initial
mass of $\approx$30-50\Msun ~
reaching the WNL stage. The evolutionary  models also predict oxygen to be 
underabundant by a factor of ten with respect to solar, and He/H up to 0.5
for evolved massive objects reaching such C and N abundances
(e.g. Maeder 1987, Maeder \& Meynet 1994, 2000, 2003). 
We thus adopted also [O]=0.1 the solar value, for
consistency with our findings of the C and N abundances, although there are no
strong oxigen lines in the range covered by our STIS spectra
for stars in this temperature range to directly measure the
oxigen abundance.

We varied He/H from 0.1 to 0.5 (by number) and in both cases we can achieve
an acceptable fit of the overall spectrum, by slightly 
tuning the other parameters.
Best fits are obtained  with \Teff = 32000~K if He/H = 0.1 
and with \Teff = 34000~K for He/H = 0.5. 
Changing the helium abundance does not only affect the strength of
the \ion{He}{2} $\lambda$ 1640, but also influences 
the relative strengths of other lines as well. 
Therefore we consider the actual helium abundance a source of uncertainty
and adopt the above range for \Teff .

Once the {\it strength} of all P Cygni lines in the spectrum was matched by the
models and the acceptable ranges constrained, we fine-tuned the trough of the 
absorptions, especially  the
\ion{Si}{4} doublet which is partly resolved. We varied 
both the terminal velocity and the rotational velocity.
While the terminal velocity significantly affects the separation
of the two absorption components of the  \ion{Si}{4} doublet,
and its uncertainty range is easily determined (see Fig. \ref{f_vinf}
and Table 2), 
rotational velocities of V$sini$ = 60 \kms and  twice as high 
V$sini$ = 120 \kms ~
were tested in the model calculations: the change only slightly affects the 
shape of the absorption troughs. 
This analysis is not as sensitive to this parameter (V$sini$) as the 
optical photospheric lines, since lines at UV wavelengths are mainly
produced in the wind thus broad and asymmetric, and the STIS spectra
resolution is too low for this purpose. 
The fit of the \ion{C}{3} and \ion{Si}{4} lines,
favours the high rotation. Again, a quantitative determination of
this parameter is not possible; we can only infer from the comparison 
an indication of possible high rotation.

All the best fit models shown are computed with log g =3.2. 
The low gravity
value was initially driven by the intention to compute models
with reasonable mass/radius/luminosity values for this type of stars.
We then tried higher gravities and obtained worse line  fits. 
 Note the good
match of both the absolute flux values (a further confirmation that our 
choice of the Radius in the best fit models is adequate) and of the slope of
the spectrum, independently confirming the temperature derived 
above by fitting the line strengths, and the extinction value
(which is also consistent with the previous work of MBHS). 
Because the spectral slope depends on \Teff 
(and reddening) and the absolute flux depends on R$^2$ and \Teff$^4$,
the overall match to both the line profiles and the absolute flux level confirms the
consistency of the parameters derived from our analysis.

The derived C/N ratio (by mass) of 0.02 is
significantly lower than the cosmic value (4.8) and in the range
of the WNL type stars (see e.g. Massey 2003).  The C/N value 
 is consistent, according to the stellar evolution calculations of
Maeder (1987), Maeder \& Meynet (1994), Meynet \& Maeder (2000, 2003) 
with the surface abundances of a massive star  
during the stage
from B supergiant to WNL.  The current mass, \Teff~ and luminosity
derived from our spectral modeling for M33-UIT236  agree with the evolutionary models
for an initial mass of about 50  \Msun .

\subsubsection{M33-UIT008}

If we superimpose the spectrum of this star to the previous one (M33-UIT236),
the two spectra, normalized to the respective continua, 
 are basically identical except for much weaker nitrogen
lines in M33-UIT008. In particular,
the \ion{C}{4} and \ion{C}{3} lines are identical, and so is the \ion{Si}{4}
doublet, except for a clearer separation of the two absorption components, 
indicating slightly  lower terminal velocity or mass-loss in M33-UIT008. 
\ion{He}{2} $\lambda$ 1640 is also identical. However, the \ion{N}{4}
and \ion{N}{5} lines are both significantly weaker than in the previous
star. We attempted to explain this difference by either a lower nitrogen abundance or 
by a lower \Teff~ or \Mdot. In our grid of models for solar CNO abundances, the 
\ion{N}{4} $\lambda$ 1720 line becomes very weak 
(comparable to the one seen in  the M33-UIT008 spectrum)
when \Teff ~ is cooler than 28000~K,  for mass-loss rates of 3-5 10$^{-6}$
\Myr , but the strength of this line also decreases  significantly with 
decreasing mass-loss rate.  
However, these models with lower \Teff ~ produce too much \ion{Si}{4} and too
little \ion{C}{3}, respect to the observed spectrum, 
if we keep the other parameters similar to M33-UIT236,
including the CNO ``WNL-type'' abundances. In short,
 a good fit of the other lines cannot be
achieved at such low \Teff .

The strength of the \ion{N}{5} doublet  significantly decreases 
in models with  lower
\Lx , however \ion{N}{4} is practically not affected by this change and
the \ion{Si}{4} doublet is significantly enhanced. Therefore, varying this 
parameter does not produce a good fit either.

Because both \ion{N}{4} and \ion{N}{5} transitions are weaker in UIT008
than in UIT236, but the spectra are otherwise identical, we
 computed  models with the same parameters as the best fit of M33-UIT236, 
but lowered the nitrogen abundance from ``WNL'' values 
down to the  solar value. This produces slightly lower (not lower enough)
nitrogen lines but enhances the \ion{Si}{4} and \ion{C}{4} emissions
enormously. Probably because we keep other parameters (such as mass-loss
rate) the same, the opacities are redistributed. 
We also tried to decrease the mass-loss rate, and alter other parameters
including different combinations of abundances, and the amount of shocks. 

The best fit, in terms of the relative line strength and match to
the flux below 1700\AA, is achieved with the parameters given in 
Table \ref{tresults}.
The results  indicate that the main difference with respect to 
M33-UIT236 is a lower mass-loss rate.

We caution that Bianchi \& Garcia (2002), Garcia \& Bianchi (2003)
encountered some difficulties in fitting weak \ion{N}{5} lines in
pop.~I Milky Way stars with the WM-basic code, and better matches could
be achieved in some cases using Hillier's CMFGEN code, although the
resulting stellar parameters were not changed (e.g. Bianchi et al. 2003a, b).
Therefore, the derived abundances, in particular of N in this case, must
be taken as indicative, as stressed in the previous section.

\subsubsection{M33-UIT104}
The  {\it line} spectrum suggests  
lower wind velocity and a lower stellar  temperature
than the previous stars examined, the ``cooler'' transitions 
(\ion{C}{3} and \ion{N}{4}) being enhanced  and the higher ions (\ion{N}{5} and 
\ion{C}{4}) weaker. A similar effect on the line strength, to some extent,
may be produced by a lower amount of X-rays from shocks.
All these effects were quantified by the modeling (Table \ref{tresults}).
A CIII line, visible as a small but detectable absorption
 in the emission of \ion{N}{5}, is the most unambiguous
 indication of \Teff~ $\leq$32000~K according to our grid of models (see Bianchi
\& Garcia 2002), at least for solar abundances. Therefore, this line
provides a strong constraint on \Teff .

In this \Teff ~ regime, the \ion{C}{3}~1175\AA~ line  becomes much more
sensitive to the C abundance than to the mass-loss rate (while \ion{C}{4}
is  saturated so rather insensitive to all parameters). This is
very fortunate because it allowed us to constrain one parameter (the C
abundance) relatively independently from the other ones in this case. 
Along the 
stellar evolution, the C-depletion develops concurrently with
the N-enrichment (according to Maeder 1987), therefore we tried different
abundance ratios for C and N, corresponding to a typical predicted
evolutionary sequence. 
The best fit is shown in figure \ref{f_wmbas}, and was 
achieved with [C]=0.1 x solar and [N]=2 x solar. 
In mass fraction, this means C/N=0.9.

 Again the match 
to both the line strength and to the absolute flux level, and the continuum shape,
is very good shortwards of 1700\AA.  

\subsubsection{M33-UIT003, M33-UIT349 and M33-UIT045}
Continuing towards progressively weaker P Cygni profiles and lower terminal
velocities,  M33-UIT003, M33-UIT045 
and  M33-UIT349 (Dec.27) 
show a smooth decrease in the lines of
 \ion{Si}{4} and \ion{C}{4}, compared to the previous targets. 
Instead,  the \ion{N}{5} $\lambda$1240
doublet  abruptly desappears in the spectra of M33-UIT045 and  of 
M33-UIT349.  In modeling the spectrum of M33-UIT045, it has been 
impossible to obtain a synthetic spectrum which matches the line features,
and the absolute flux, with nitrogen abundance solar or higher. 
The best fit spectrum (shown in Figure \ref{f_wmbas}) was obtained by lowering
the nitrogen abundance by a factor of ten with respect to solar.
A nitrogen underabundance in such evolved massive star is hard to explain
and extremely suspicious.
Again, we caution that model calculations may be less reliable for such
extreme parameters than for the previous cases.

Similarly to M33-UIT045, it has been extremely difficult to produce a synthetic
spectrum without \ion{N}{5} to match the observed spectrum of M33-UIT349.
The fit  was achieved with an extremely low mass
loss rate, about two order of magnitude lower than the other stars in the sample.
The result is consistent with the fact  that the typical wind lines in the UV
spectrum are almost absent. However, the  low mass-loss rate is
 surprising since this star
has photospheric parameters not too different from the rest of the sample. 
We must keep in mind that additional faint sources may contribute 
-although marginally - to the spectrum,
but they are not resolved nor confirmed.
It is also puzzling that the wind lines in the  
 M33-UIT349 spectrum seems {\bf wider}
rather than narrower, compared to the spectra with stronger lines.
Typically, a progression from stronger to weaker P Cygni profiles (i.e. decreasing
 mass-loss) also corresponds to decreasing wind velocities, as it evident
e.g. in the sequence  (from the top down) of spectra in 
Figures \ref{f_spectra_all} and \ref{f_vinf}.  
The large velocity may suggest a physical binary, i.e. 
two stars with high ($\approx$ 1000\kms ) velocity relative to each other,
both having weak lines. Or, more simply, the fact that 
the lines are so weak (and the limited S/N) prevents to resolve blends
and makes the analysis inconclusive in this case.

Finally, for  two  ``weak-lined'' spectra, modeling was achieved
with a carbon underabundance (respect to solar) of   a factor of two,
and not twenty like the hotter stars with stronger winds, and
lower mass values. 
The derived parameters \Teff , \lbol, Mass, are again 
 consistent with the evolutionary models of Maeder, since lower
mass stars produce less enhancement of N and less depletion of C,
than more massive stars. 

Looking at the absolute flux level and slope at longer wavelengths (Figure 7), 
we note a slight mismatch between model and observed spectrum 
for M33-UIT008, M33-UIT045 and M33-UIT003. 
While the best fit model reproduces well the continuum and  all the lines
in the shorter wavelengths range (where the flux from the
hot star dominates), 
the slope of the entire observed spectrum cannot be matched perfectly by  any
single star model for any combination of reddening (using known reddening types).
The mismatch between model and observed spectra, however, 
is of the order of 0.1 - 0.2 
10$^{-14}$ergs cm$^{-2}$ s$^{-1}$, 
or $\approx$ up to 5\% beyond 
$\approx$ 1500\AA, thus comparable to the calibration uncertainty.
The flux level and the line spectrum are matched well in the short wavelength
range, pointing to a consistent and plausible model fit for the hot star.
In the finding chart produced from the STIS CCD acquisition
image 
and the spectral intensity cuts shown in 
Figure \ref{f_ycut}, no extra faint components can be clearly resolved.
Therefore, the mismatch could indicate either an unresolved chance superposition
of a fainter cooler star (statistically unlikely), 
or a faint binary component with separation
less than $\approx$ 0.2pc (unresolved), or an  UV  extinction slightly different
than the types known so far in the MW or LMC (see next section).
The level of mismatch is however of the order of the calibration accuracy
(see Section 2.2.), therefore it could just be an instrumental effect.

\section{Discussion and Conclusions}

We analyzed STIS Ultraviolet spectra of six Ofpe/WN9 stars
in M33.  Previous optical spectra of our sample (MBHS) showed a
 progression from
``weak-lined'' slash stars, similar to BE470 in the LMC
(with strong P Cygni HeI and Balmer emission but very weak NIII
$\lambda$ 4634,42 and HeII $\lambda$ 4686) to ``strong lined"
slash stars, similar to BE381 in the LMC (with strong NIII and HeII).
The UV line spectra show a progression (Figure 1)  
of line intensities which does not correlate
with the earlier optical spectra (Figure 2), suggesting possible  
 variability  on  short time-scale
(years) of the mass-loss rate.

\begin{figure}
\plotone{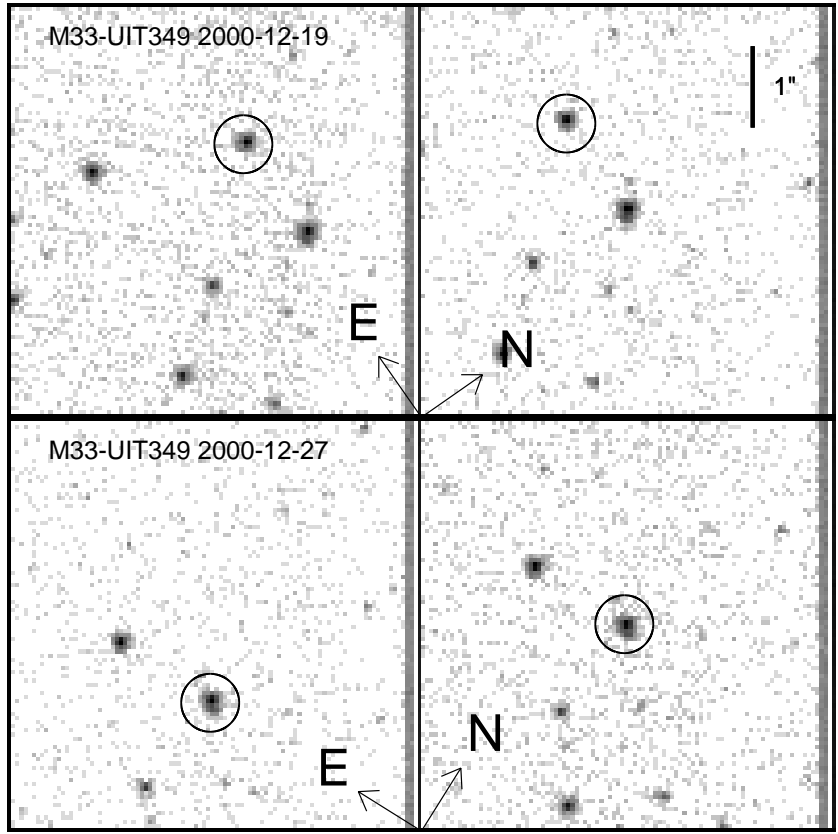}
\caption{Acquisition images of M33-UIT349. Left: before centering - 
Right: after coarse acquisition centering. 
Note that the orientation is  different on the two dates
(PA=54.65 on Dec.19 and PA=31.64 on Dec.27).
The target was acquired and perfectly centered in the coarse stage, 
on both dates, however on Dec. 19 during  the 
fine acquisition there was a jump and another star of equal brightness
(in the circle) was put in the slit.
\label{f_acq_349} }
\end{figure}

\begin{figure}
\plotone{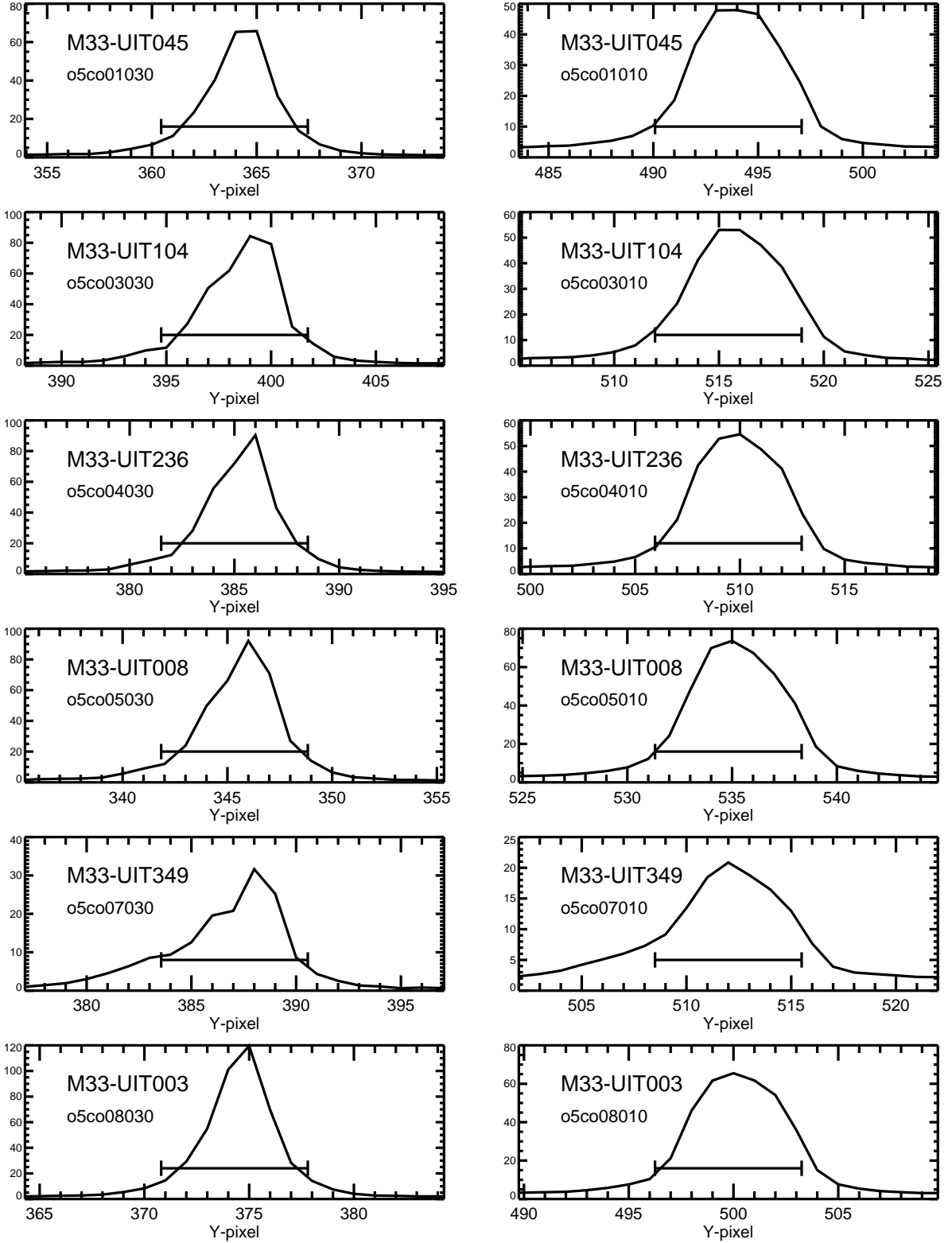}
\caption{Intensity profile in the spatial direction for the G140L 
(left) and G230L (right) spectra.
 for all the program stars. The profile is the sum of 301 columns around the center
of the image.
The data shown for M33-UIT349 are the Dec.27 observations, when
the correct target was observed. 
\label{f_ycut} }
\end{figure}

\begin{figure}
\plotone{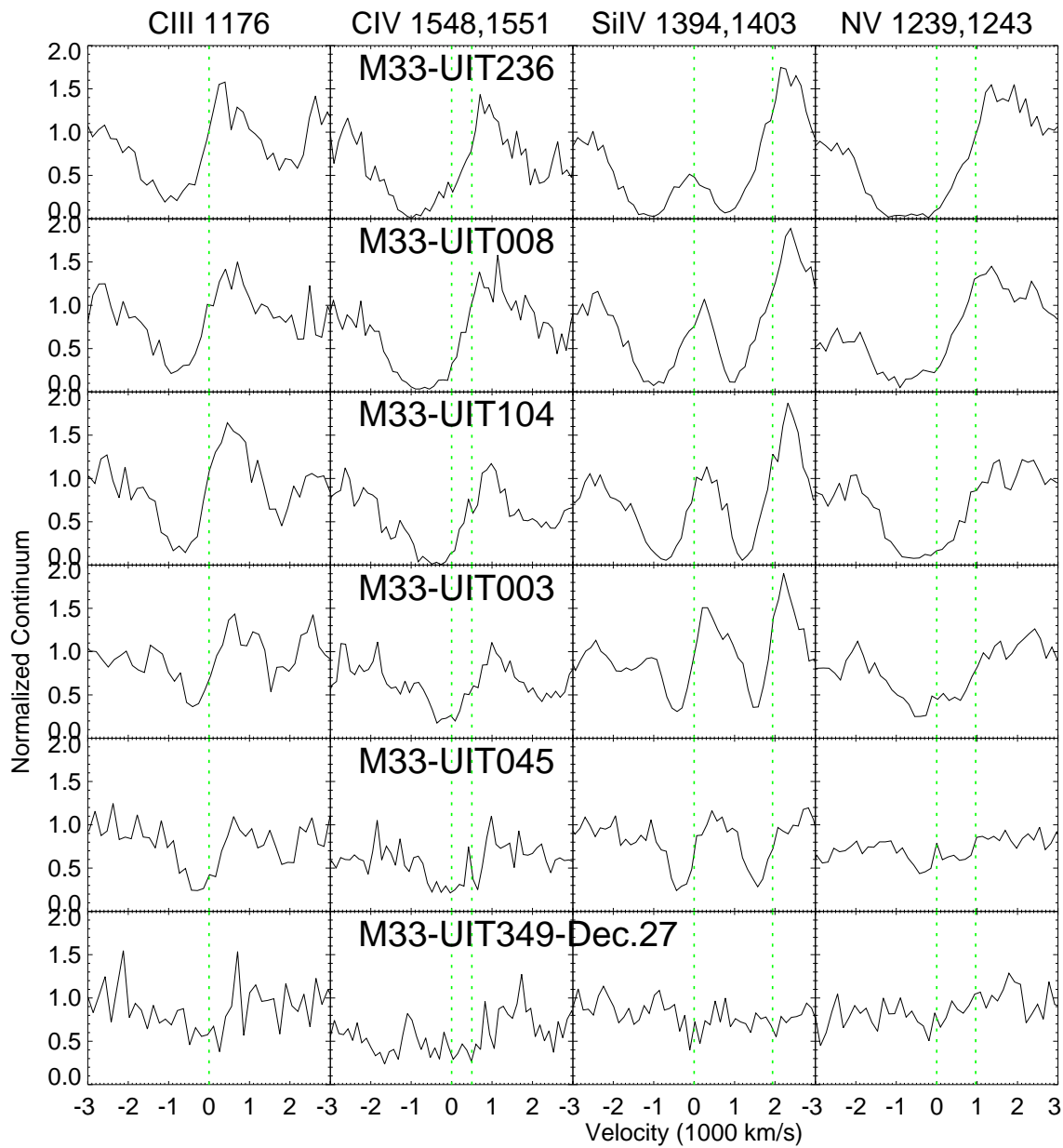} 
\caption{The four  
 strongest UV wind lines are shown in velocity scale 
(thousand \kms) for the Ofpe/WN9 program stars. 
Vertical green (dotted) lines indicate the rest positions of the transitions.
The recession velocity of each star (Table \ref{tdata} ) has been
removed. From top
to bottom, the lines show a progression of decreasing strength and
width (i.e. decreasing wind terminal velocity). The effect is especially
evident in the \ion{Si}{4} doublet, where the two components are more
separated in wavelength (while the \ion{C}{4} and \ion{N}{5} doublets
are always blended in velocity).
\label{f_vinf} }
\end{figure}

\begin{figure}
\plotone{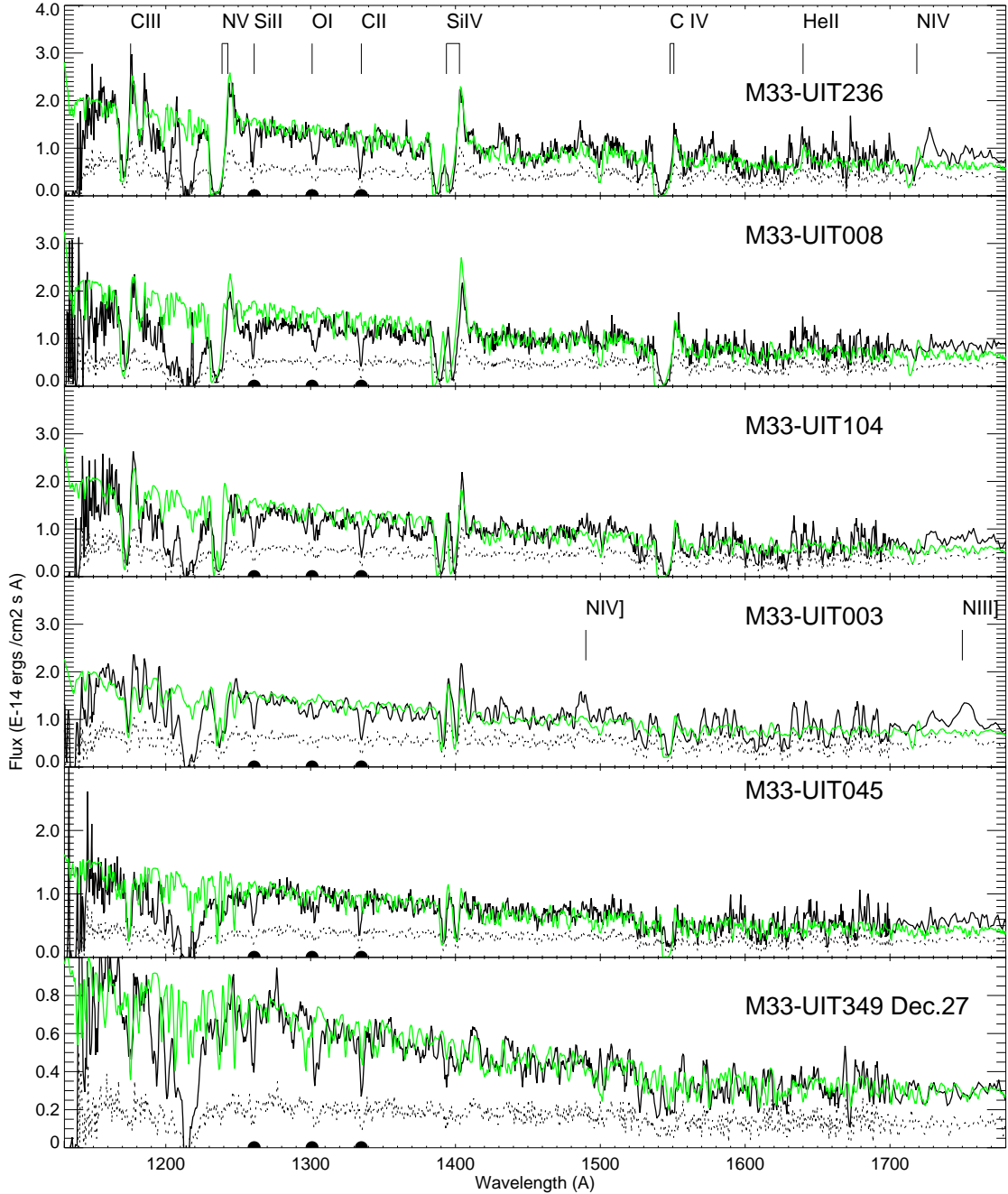}
\caption{The STIS spectra (black line), dereddened for the amount of
extinction given in Table 2, and the best fit models (green/light grey) computed with 
WM-basic. The model parameters are compiled in Table 2. 
The dotted line is the observed spectrum with no de-reddening, to illustrate
the sensitivity of our modeling to even such small reddening amounts. 
The range below
1800\AA ~ is shown, where all the strong stellar wind lines are.
The strongest interstellar lines in this range,
\ion{Si}{2}$\lambda$1260, \ion{O}{1}$\lambda$1301, and \ion{C}{2}$\lambda$1335,
are marked with a filled bubble at the bottom of the spectra.  
In the spectrum  of M33-UIT003 
two broad emission features are seen, near 1490\AA~ and 1750\AA. 
There are a number of NIV] and NIII] intercombination lines around
those wavelengths respectively. The emission is probably not of stellar origin.
\label{f_wmbas} }
\end{figure}

 One target, M33-UIT349, was observed twice, 8 days apart, but
successfully only on Dec. 27.  Its spectrum seems to include a very faint extension
 (Figure \ref{f_ycut}), separated spatially by $<$0.1''
(thus if physically associated, by $<$ 0.4pc), but
the source is not resolved into separate components at the STIS
CCD resolution (Figure \ref{f_acq_349}).
 In any case, the apparent
 secondary component has negligible flux with respect to the main source
(figure \ref{f_ycut}), as confirmed by the good fit to the spectral
distribution.  
The spectrum shows
extremely weak 
wind lines.

 For three stars (half of the sample) we find  C/N=0.02
by mass, consistent with predictions from stellar evolution models
for an evolved massive star 
(initial mass $\approx$40-50 \Msun) in the phase transitioning from
Blue supergiant to WNL 
(i.e. towards the W-R stage) at constant luminosity in the H-R diagram.
The other physical parameters (\Teff, mass, luminosity) are
also entirely  consistent with the 
 scenario of massive stars approaching the WNL stage
(Maeder 1987, Maeder \&  Meynet 1994, Meynet \& Maeder 2000, 2003). 
It must be kept in mind that the available spectroscopic range provides
several wind lines but no photospheric lines, therefore derivation of
abundances is not very accurate as wind line strengths depend also
on the mass-loss rate and other factors. However, the consistency of
the entire modeling (lines, absolute flux, velocity) supports the results
as quite reliable,  and all the parameters indicate a consistent evolutionary
 picture. 
For the other three stars, C was found to be (by number), only
about half the solar value. For M33-UIT104, the analysis  indicates N to be 
overabundant by a factor of two, i.e. C/N = 0.9 by mass.
For the other two stars with extremely weak lines, the nitrogen abundances
and the overall modeling is way more uncertain, for the reasons explained in 
the previous section.
Either these stars are in an earlier evolutionary stage,
 evolving  off the ZAMS, still towards 
lower \Teff 's, or their progenitors has lower initial masses
(by comparison with Maeder's evolutionary calculations).
Variability of the mass loss rate is also  
 possibile, as mentioned, complicating the interpretation.

Studies of coeval clusters in the Milky Way (Massey, DeGioia-Eastwood, \&
Waterhouse 2001) and Magellanic Clouds (Massey, Waterhouse, \& DeGioia-Eastwood
2000) suggest that LBVs are descendent from the most massive stars ($>90\cal
M_\odot$).  The two Ofpe/WN9 stars in their sample, however,
come from stars of much lower mass, $>$ 25-35$\cal M_\odot$, calling into
question the evolutionary link between LBVs and Ofpe/WN9s. 
The range of progenitor masses inferred by comparing  our derived current
masses, abundances, \Teff~ and luminosities with evolutionary 
calculations ($\approx$ 30-50\Msun), 
is consistent with their findings.
On the other hand,  a link between LBVs and Ofpe/WN9 has
been observationally established in the cases e.g. of AG Car, R127, S61, S119
(Stahl 1987, Stahl et al. 1983, Pasquali et al. 1999, Nota et al. 1995)
although  these studies do not provide enough statistics
to  quantify the correlation in terms of evolution. 
Langer et al. (1998) and Lamers et al. (2001) discuss the effect of stellar rotation,
on the evolution of massive stars,
and  show that rotation "extends" the occurrance of the LBV phase to
lower progenitor masses.

One important parameter, the mass-loss rate, varies from 8 10$^{-6}$ \Myr to
2 10$^{-6}$ \myr among the first four stars in table \ref{tresults}, which
appear to have similar masses (within the uncertainty of our analysis) but
a range of temperatures. The uncertanties on the mass-loss rates are
less than 30\% according to our model calculations. 
Although the difference in \Teff ~ within our sample 
covers a fairly narrow range, 
and the sample is numerically limited, there is 
a trend  of mass-loss increasing as the star evolves towards the
W-R stage and becomes hotter. Mass-loss rate 
increases with luminosity (e.g. Bianchi \& Garcia 2002, Nugis \& Lamers 2000,
Vink et al. 2001), but the small range of luminosity variation among the
sample does not explain the large mass-loss rate spread (see below). 
 
The values of \Mdot ~ that we derive 
for the ``strong-lines'' stars (the first four in Table \ref{tresults}) 
are lower than the high mass-loss rates
of W-R stars, and higher than mass-loss rates expected for O stars with similar parameters.
 Maeder's evolutionary  models assume an average 
\Mdot = 4 10$^{-5}$ \Myr for WNL stars, and the  Nieuwenhuijzen \& de Jager (1990)
parametrization for the earlier stages (this parametrization is
based on a compilation of observed mass loss rates).  Our mass-loss rates  are
 also lower than Nieuwenhuijzen \& de Jager (1990)
parametrization which, for the physical parameters given in table \ref{tresults},
would predict for the first four stars of the sample \Mdot~ between 2.1 and 1.5 
10$^{-5}$ \Myr. 
Recent works revised the mass-loss rates for Wolf-Rayet stars, taking into
account ``clumping'' in the wind. These 
mass-loss rates are generally lower than earlier determinations,  of 
the order of $\ge$ 1 10$^{-5}$ \Myr for WN-type stars of comparable luminosities
to our targets (Nugis \& Lamers 2000; Nugis, Crowther \& Willis 1998; 
Graefner, Koesterke \& Hamann 2002), thus higher than what we measure for
the Ofpe/WN9 stars.
To compare with O-type stars,  we use the  recent ``recipe'' 
(based on theoretical monte-carlo simulations) from Vink et al. (2000, 2001),
for pop.I stars,  and find that the predicted 
mass-loss rates are of the order of 2 10$^{-6}$ \Myr for the first five
stars (2.4 to 1.7 10$^{-6}$ \Myr from M33-UIT236 to M33-UIT045), for solar metallicity,
and less for metallicity lower than solar (e.g. for z=0.1 $\times$  solar, 
\mdot ~ would be lower by a factor of seven). 
In summary: the values of mass-loss rates derived for 
our sample  are lower than typical W-R mass-loss rates, but
in comparison to predictions
from  radiation-pressure wind theory for pop.I stars,
mass loss is enhanced  for the hottest stars
(M33-UIT236, M33-UIT008, M33-104), is comparable to the predictions for M33-UIT003,
and lower than predicted for M33-UIT045 and M33-UIT349. However, as we stressed before,
 \Mdot ~ derivations are  less reliable for the two latter stars, as the
lines are extremely weak. 
Another result emerging by these comparisons is that 
mass-loss rate varies within the sample more than we would expect from the
variation of the other physical parameters (\Teff , \Lbol, mass).  For this fact, and 
because the UV line strengths do not correlate with optical line strengths
observed several years earlier, we suggest that mass-loss rate may vary in time as well. 
LBV stars also display
variable mass-loss rates, which Vink \& de Koter (2002) 
explain in terms of   changes in the line driving efficiency. 
The large spread in mass-loss rates among our sample may also
relate to evolutionary effects. 
The physical conditions of the radiation-pressure driven wind change during
the evolutionary phases from O-type to LBV to W-R, and these classes
of objects display very different mass loss rates and wind velocities,
although the stellar luminosity remains basically constant during the
transition.  According to Lamers \& Nugis (2002), the wind changes
are mainly due to change in stellar parameters (radius, gravity) and
to a lesser extent to change in surface chemical abundances. 
 The ensemble of our findings
consistently points out that the first four objects 
in Table \ref{tresults} are transitioning towards
the W-R (WNL) stage through an enhanced mass-loss phase.

All spectra are very well fitted in both continuum flux level
and line strength at the short wavelengths (G140L range), while a  mismatch
between observed flux and model   is seen longwards for most stars.
All stars appear to be 
point-like sources at the STIS spatial resolution. 
The amount of the mismatch (up to 5\% at the longer wavelengths)
 is comparable  to the absolute flux
confidence level (section 2.2), 
and to the uncertainty in the extinction amount and
extinction law.  A variation in E(B-V) of 0.01 would have an effect 
comparable to the mismatch seen beyond 1700\AA.  However, it would create
a larger mismatch at the shorter wavelength, which are more sensitive to 
the reddening, thus was excluded. 
The amount of extinction (\ebv ) adopted as our best
fit to the spectra 
is given in table \ref{tresults}, and the adopted
extinction law is a combination of Galactic extinction (foreground)
and LMC-type extinction.  
This was found to be generally the preferrable solution, after 
examining different combinations with all known extinction laws
(for other galaxies), concurrently with varying the stellar model
parameters. The UV extinction law in M~33 is currently being investigated
by us with another STIS program, and the results may refine this issue.
We  attempted to add different 
black-body components to the stellar-model fits. While we may visually
improve the fit to the observed spectra at the long wavelengths,
removing the mismatch,  
the results would not be significant
given that the flux level of a postulated additional component 
would be  up to a few \% that
of the luminous hot star. 
Thefore, we cannot conclude at this point,  whether
 a faint excess flux, peaking at about 1800\AA ~ or longwards,
 is present in  addition
to the brighter  hot luminous star spectrum, or if the extinction curve
slightly differs from the MW and LMC ones.  

Finally, we point out   again two remaining possible sources of uncertainty 
in the present analysis. The first  may come from some inconsistency in the 
WM-basic calculations,  for the cases of extreme parameters such as 
those of M33-UIT045
and M33-UIT349, where the flux indicates a hot, luminous object but
the lack of wind lines (or just of NV in the case of M33-UIT045) indicates
an unusually low mass-loss rate. These  caveats are
discussed at length by Bianchi \& Garcia (2002,2003b), Garcia \& Bianchi (2003). 
Second, the amount of shocks in the winds 
and the effect of the related X-rays on the ionization - 
which bares on the \Teff ~ results, 
cannot be precisely estimated when only the UV range is available. The
FUSE range, at shorter wavelengths, contains transitions much more sensitive to
this parameter than those in our STIS spectra, primarily the \ion{O}{6} doublet
(Bianchi \& Garcia 2002, Bianchi et al. 2003). We plan to extend the
analysis of the sample to the FUSE range as a next step, although the fluxes
of these distant objects are at the limit of detection for FUSE.

It is  interesting to compare our findings with similar objects
in other galaxies, to verify metallicity effects on the evolution. 
To our knowledge, no similar modeling of UV spectra has been performed
for Ofpe/WN9 stars in the MW or other galaxies. However, previous
works, mostly based on optical spectroscopy, provide very useful comparisons. 
Similarly to our STIS spectra, 
UV FOS spectra of all the LMC ``slash stars"
 show P Cygni profiles  of
CIII $\lambda$ 1176, SiIV$\lambda$$\lambda$ 1394,1403 and
NIII $\lambda$$\lambda$ 1748, 1752.
Other strong lines,  CIV$\lambda$$\lambda$ 1548,1550,
 NV $\lambda$$\lambda$1238,1243, HeII $\lambda$ 1640, NIV$\lambda$1719 and
AlIII $\lambda$$\lambda$ 1855,1863, vary from pure absorption to
strong P Cygni profiles in the sample (Pasquali et al.
1997). 
these authors derive terminal velocities of the order of 400
km s$^{-1}$ ,
much lower than the wind velocities  measured in our sample,
and mass-loss rates (from H$\alpha$ equivalent widths) 
 of $\approx$ 2 - 5 10$^{-5}$
M$_{\odot}$ yr$^{-1}$, much higher than what we found  for the M33 counterparts.
The discrepancy may be due to the H$\alpha$ equivalent width 
method, where the correction for the underlying photospheric absorption is a
very large source of uncertainty (Bianchi \& Scuderi 1999).
A non-LTE analysis of the  optical spectrum of an Ofpe/WN9 star 
in NGC~300 was performed  by Bresolin et
al. (2001) using the  Hiller \& Miller (1998) code. The \Teff ~ is lower, but the
luminosiy and the mass loss rate higher, than the range of values 
found in our sample. 
Crowther et al. (1977)  performed a non-LTE analysis of optical spectra
of LMC's Ofpe/WN9 stars, which they however re-classified as WNL (WN9-11). 
They find a range
of  \Teff ~ which overlaps with the lower end of our range,  most of their sample
(WN9-11) stars having lower \Teff's than ours,
and higher \Mdot  ~ values.  We caution again that mass loss rates derived from UV and 
from optical spectra give often discrepant results (e.g. Crowther et al. 2002,
Bianchi \& Garcia 2002), thus the comparison has to be taken with caution.
A more conclusive comparison would require similar analysis of similar data-sets.  

\begin{acknowledgments}
Support for this work (proposal GO~8207) was provided by NASA through 
 grant  GO8207-0197A from
the Space Telescope Science Institute, which is operated by the Association
of Universities for Research in Astronomy, Inc., under NASA contract
NAS5-26555. We are very grateful to Miriam Garcia for computing several
WM-basic models of the grid used in this work, and to Jorick Vink,  Lars Koesterke and
Wolf-Rainer Hamann for useful discussions.   
\end{acknowledgments}

\newpage

\newpage

\newpage


\newpage

\newpage
  
\newpage

\newpage
\begin{figure}
\plotone{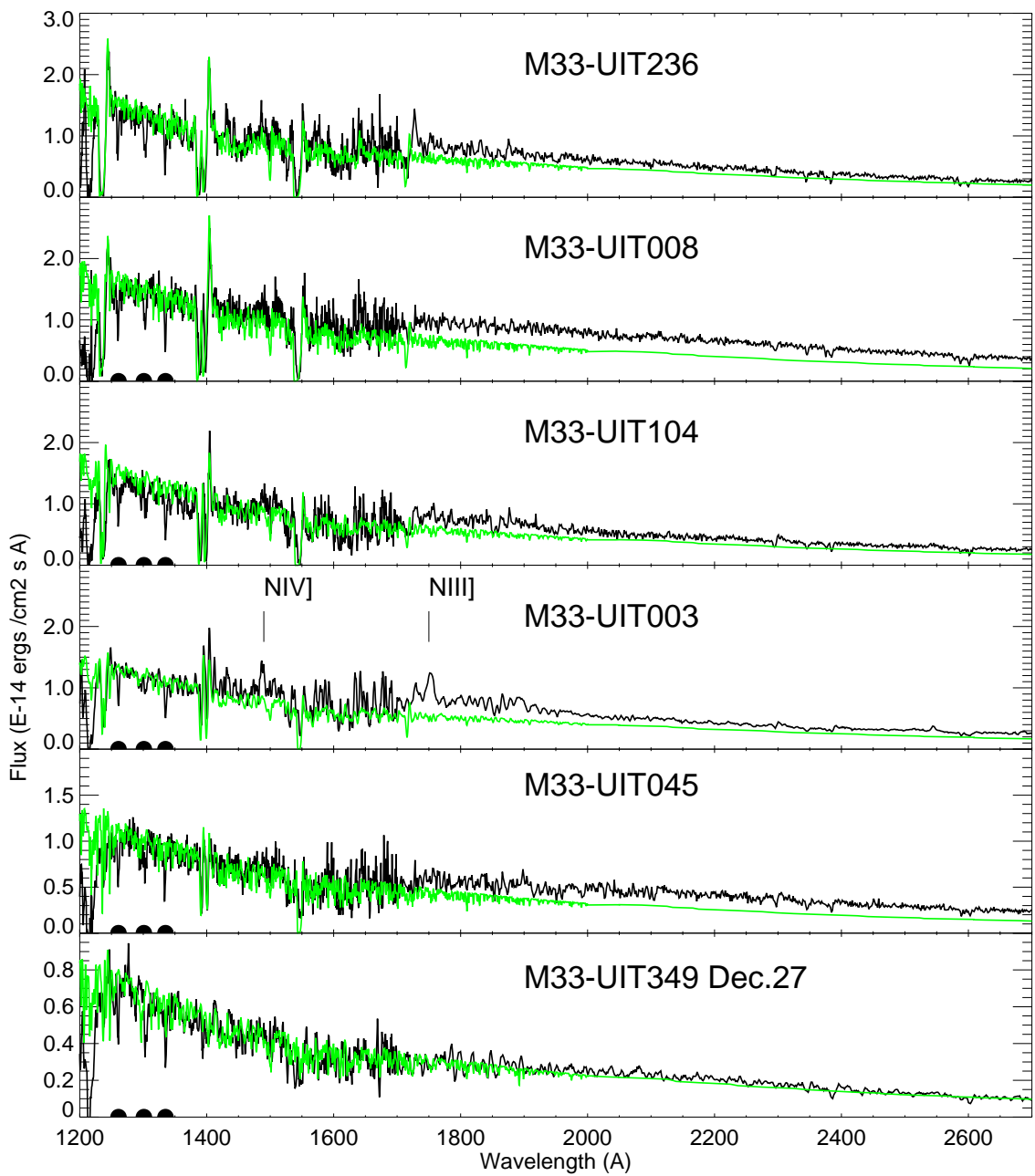}
\caption{Same as in Figure 6, but the longer wavelengths range is shown.
The STIS spectra (black line), dereddened for the amount of
extinction given in Table 2, and the best fit models computed with 
WM-basic. 
A slight flux excess is seen in most spectra beyond $\approx$ 1800\AA,
its possible origin is discussed in the text.
\label{f_wmbas_spare} }
\end{figure}

\newpage

\renewcommand{\arraystretch}{.6}
{\footnotesize
\rotate
\begin{deluxetable}{lccclcccc}
\tablewidth{0pt}
\tablecaption{The program stars and the STIS data {\label{tdata}} }
\tablehead{
\multicolumn{1}{l}{Star name}  & \multicolumn{1}{c}{R.A.}    & 
\multicolumn{1}{c}{Dec.}       & \multicolumn{1}{c}{D$_Gal$} &
 \multicolumn{1}{c}{V}         & B-V                         & 
 U-B                           & \multicolumn{1}{c}{STIS Data-sets} & 
\multicolumn{1}{c}{M33 velocity} \\
\multicolumn{1}{l}{ }          &                  (2000)     & 
  (2000)                       & \multicolumn{1}{c}{ [Kpc]}  & 
                               &                             & 
                               & \multicolumn{1}{c}{G230L,G140L}      
                               &\multicolumn{1}{c}{ [\kms ] }  
}
\startdata
\multicolumn{5}{l}{\it The Ofpe/WN9 stars\rm } \\
M33-UIT236 & 01 33 53.58 & 30 38 51.8 & 0.4 &18.08 &-0.14 & -0.84 & O5CO04010,20,30  &-165.7 \\
M33-UIT008 & 01 32 45.38 & 30 38 58.6 & 6.0 &17.67 &-0.07 & -1.01 & O5CO05010,20,30  &-145.4 \\
M33-UIT104 & 01 33 27.22 & 30 39 09.1 & 2.1 &18.03 &-0.14 & -0.91 & O5CO03010,20,30 &-147.5 \\
M33-UIT003 & 01 32 37.70 & 30 40 05.7 & 6.8 &17.44 & 0.01 & -0.98 & O5CO08010,20,30 &-151.1 \\
M33-UIT349 & 01 34 18.66 & 30 34 11.6 & 3.5 &18.71 & 0.57 & -0.83 & O5CO07010,20,30 &-153.0 \\
M33-UIT045 & 01 33 09.10 & 30 49 54.5 & 5.8 &18.02 & 0.00 & -1.01 & O5CO01010,20,30 &-215.9 \\
           &             &            &     &      &      &       &             &       \\
\multicolumn{5}{l}{\it Serendipitous sources\rm } \\
M33-UIT349-B & 01 34 19.66 & 30 34 11.6 & 3.5 & ...& ...  & ...   & O5CO06010,20,30 &-153.0 \\
M33-UIT339 & 01 34 16.06 & 30 36 42.3 & 2.8 &18.03 & 0.01 & -0.83 & O5CO02010,20,30 &-173.4 \\
M33-UIT340 & 01 34 16.04 & 30 36 38.0 & 2.8 &18.42 & -0.04& -0.81 & O5CO02010,20,30 &-173.4 \\
\enddata
\end{deluxetable}
}

\newpage
\normalsize
{\small
\renewcommand{\arraystretch}{.6}
\rotate
\begin{deluxetable}{lcccclccccccc}
\tablewidth{0pt}
\tablecaption{The derived stellar parameters  {\label{tresults}} }
\tabletypesize{\scriptsize}
\tablehead{
\multicolumn{1}{l}{Star name}      & \multicolumn{1}{c}{E(B-V)} & \multicolumn{1}{c}{\Vinf } & 
\multicolumn{1}{c}{T${_eff}$}      & \multicolumn{1}{c}{log g}   &
\multicolumn{1}{c}{L/L$_{\odot}$}  &  R/R$_{\odot}$              & 
 \multicolumn{1}{c}{\Mdot}         & \multicolumn{1}{c}{\Lx }    & \multicolumn{1}{c}{M} & 
\multicolumn{1}{c}{[C]/[C$_\odot$] }       & \multicolumn{1}{c}{[N]/[N$_\odot$]}       &  C/N \\
\multicolumn{1}{l}{ }          &         [ mag ]                &  \multicolumn{1}{c}{\kms}   &   
      [ K ]                    & \multicolumn{1}{c}{ }           &  &  
     &  [\myr]                   & \multicolumn{1}{c}{}   &   
\multicolumn{1}{c}{[\Msun]}    &\multicolumn{2}{c}{ } & (by mass) 
}
\startdata 
M33-UIT236       & 0.10 & 1950$\pm$150&33,000$\pm$1500 & 3.2 & 5.67 & 21 & 8.e-6 & -7.5& 26 &0.05&10. & 0.02\\
M33-UIT008       & 0.08 & 1950$\pm$150&32,000$\pm$1000 & 3.2 & 5.66 & 22 & 4.e-6 &-7.75& 28 &0.05&10. & 0.02\\
M33-UIT104       & 0.09 & 1600$\pm$150&31,000$\pm$1500 & 3.2 & 5.61 & 22 & 3e-6 &-8.0 & 28 &0.50& 2. & 0.90\\
M33-UIT003       & 0.07 & 1000$\pm$120&30,000$\pm$1500 & 3.2 & 5.55 & 22.& 2e-6 &-10. & 28 &0.05&10. & 0.02\\
M33-UIT045       & 0.10 & 950$\pm$100 &30,000$\pm$1500 & 3.2 & 5.47 & 20 & 5e-7 &-9.5 & 23 &0.5 &0.1:: & 17::\\
M33-UIT349       & 0.11 & 1700: &27,000$\pm$2000:      & 3.0:& 5.28:& 20:& 1e-8:&-7.0:& 15:&0.5:&2.: & 0.90\\
\enddata
\end{deluxetable}
}
\normalsize
\end{document}